\documentclass[aps,prl,preprintnumbers,amsmath,amssymb,latexsym,nofootinbib,array,enumerate,letter,twocolumn,superscriptaddress]{revtex4-2}

\usepackage{here}
\usepackage{comment,braket}
\usepackage{epsf}
\usepackage{amsmath}
\usepackage{graphics}
\usepackage{amsfonts}
\usepackage{amssymb}
\usepackage{latexsym}
\usepackage{color}
\usepackage{ulem}
\usepackage{feynmf}
\usepackage{enumerate}
\usepackage{lipsum}
\usepackage[pdftex]{graphicx}
\usepackage{bm}
\usepackage[pdftex]{hyperref}
\usepackage[svgnames]{xcolor}
\definecolor{phthaloblue}{rgb}{0.0, 0.06, 0.54}
\hypersetup{
    colorlinks=true,
    linkcolor=phthaloblue,
    citecolor=blue,
    filecolor=blue,
    urlcolor=phthaloblue,
    }

\newcommand{\GeV}{\  {\rm GeV} }
\newcommand{\TeV}{\  {\rm TeV} }
\newcommand{\pr}[1]{\left(#1\right)}

\newcommand{\micromegas}{\code{MicrOMEGAs5.0}~}

\newcommand{\code}[1]{\texttt{#1}}

\begin{document}
    
\title{
Discovery potential for split supersymmetry with thermal dark matter
}

\author{Raymond T. Co}
\affiliation{William I. Fine Theoretical Physics Institute, School of Physics and Astronomy, University of Minnesota, Minneapolis, MN 55455, USA}
\author{Aaron Pierce}
\affiliation{Leinweber Center for Theoretical Physics, University of Michigan, Ann Arbor, MI 48109, USA}
\author{Benjamin Sheff}
\affiliation{Leinweber Center for Theoretical Physics, University of Michigan, Ann Arbor, MI 48109, USA}
\author{James D. Wells}
\affiliation{Leinweber Center for Theoretical Physics, University of Michigan, Ann Arbor, MI 48109, USA}
\date{\today}

\begin{abstract}
Supersymmetric extensions of the Standard Model with scalar superpartners above 10 TeV are well motivated since the Higgs boson mass can be explained by quantum corrections 
while maintaining gauge coupling unification. If supersymmetry breaking is transmitted to gauginos via anomaly mediation, the gaugino masses are loop suppressed compared to scalar masses, and the lightest supersymmetric particle is the Higgsino or wino, which can be the dark matter. In this setup, we identify the regions of parameter space that reproduce the observed Higgs boson mass and the thermal abundance of dark matter. We analyze the effects of complex phases in the gaugino mass parameters on the electron electric dipole moment (EDM) and the dark matter scattering cross section. We find that, for scalar masses up to $10$ PeV and any size of the complex phases, the model with Higgsino dark matter is  within reach of planned experiments---Advanced ACME via electron EDM and LUX-ZEPLIN via dark matter direct detection---with complementary discovery potentials, and the model with wino dark matter is within reach of future electron EDM experiments.
\end{abstract}

\preprint{LCTP-22-09, UMN-TH-4126/22, FTPI-MINN-22/17}

\maketitle

\section{Introduction}

Supersymmetry (SUSY) remains a promising extension to the Standard Model of particle physics. It provides for unification of the gauge couplings \cite{Dimopoulos:1981yj}, and the lightest superpartner can provide an explanation for the abundance of dark matter (DM) \cite{Goldberg:1983nd,Ellis:1983ew} if stabilized via $R$-parity. Pressure from a number of experiments, including colliders (see reviews~\cite{Alves_Simplified_2012,Canepa_Searches_2019,Baer:2020kwz}),  Flavor Changing Neutral Current (FCNC) experiments (see Refs.~\cite{Donoghue_Flavor_1983, Hall:1985dx, Baer_Landscape_2019} or a review~\cite{Gabbiani:1996hi}), and  proton decay searches
\cite{Weinberg_Supersymmetry_1982, Sakai_Proton_1981,  Chamoun_Nucleon_2020}, tends to push sfermion masses well above the TeV scale.  
While such heavy scalars strain traditional notions of naturalness, there is no firm empirical reason to exclude this possibility (further discussion on naturalness can be found in~\cite{Dijkstra_Naturalness_2019,Giudice:2008bi,Wells:2021zdp}).

In the minimal supersymmetric standard model (MSSM), the lightest Higgs boson has mass less than the $Z$ boson at tree level. 
However, one-loop corrections can be significant~\cite{Okada:1990vk,Ellis:1990nz,Haber:1990aw,Okada:1990gg} and scale with the logarithm of the separation of the SM and the SUSY scales.  The observed value of the Higgs boson mass may be explained if the stop squarks are heavy, with masses 
of order $10-100$ TeV.
One possibility is that the scalars receive a large, soft supersymmetry breaking mass of this size~\cite{Arvanitaki:2012ps}, but the gaugino masses are suppressed, as can occur if there is no singlet in the SUSY-breaking sector.  In this case an anomaly-mediated contribution to the gauginos is dominant~\cite{hitoshi_gaugino_anomaly_1998, randall_out_1999,Gherghetta:1999sw}. This generates gaugino masses  suppressed by a loop factor relative to gravitino mass, which we take to be equal to a universal scalar masses, $m_{0} \simeq m_{3/2}$.\footnote{Anomaly mediation also generates one-loop suppressed contributions to scalar trilinear couplings as well as two-loop suppressed contributions to scalar (mass)$^2$.  The former have negligible effects in the scenarios considered here, and the latter are subdominant to the universal piece considered above, though both are included in this analysis.}

The anomaly-mediated contributions to gaugino masses are proportional to the relevant gauge beta function, with numerical values given by
\begin{equation}
    M_3 \approx 3 M_1 \approx 10 M_2 \approx m_{0} / 30,
\end{equation}
with the wino mass being the lightest at $M_2\sim m_{0}/300$, and $M_{1,3}$ being the bino and gluino masses respectively. 

Under the above assumptions, we can specify the MSSM spectrum with four parameters, the universal scalar mass $m_0$, the ratio of the Higgs boson vacuum expectation values $\tan\beta$, the Higgsino mass parameter $\mu$, and the associated soft term $B\mu$.
We are agnostic to the origin of the $\mu$ term and the $B \mu$ term.   Generically, in a pure anomaly mediation setup, we might expect that $B \mu$ would receive a contribution $\mathcal{O}(m_{0} \mu)$ from the conformal anomaly, but here we will allow the possibility of additional contributions to this term, and as discussed below, we will fix this term as well as $\tan \beta$ by the requirement of proper electroweak (EW) symmetry breaking.

In this setup, depending on the value of $\mu$, either the Higgsino or wino will be the lightest superpartner. As these are both electrically neutral and weakly interacting, they are candidates for DM that can be produced via thermal freeze-out in the early universe
~\cite{Lee:1977ua, Steigman:1984ac}. In order for the lightest superpartner to form the entirety of the known DM relic density, it must have mass $m_\chi = 1.1 \pm 0.2$ TeV in the case of Higgsino DM~\cite{Profumo_Statistical_2004,giudice_split_2005,Pierce:2004mk,Hryczuk:2010zi}, or  $m_\chi = 2.8 \pm 0.2$ TeV for wino DM~\cite{Hisano:2006nn,Cirelli:2007xd,Hryczuk:2010zi}, implying $\mu = 1.1$ TeV or $m_0 \simeq 1.2$ PeV, respectively.

We will explore the model we have described above, namely split SUSY with universal scalar masses equal to the gravitino mass and with gaugino masses and trilinear couplings specified by their anomaly-mediated values.  Within this framework we find the regions of parameter space with thermally produced Higgsino or wino DM. We assume that the unified scalar masses 
are specified at the GUT scale 
and that the Higgs sector parameters $\mu$ and $B\mu$ take on values that meet the requirements for EW symmetry breaking. The free parameters will then be the scalar mass $m_0$ and $\tan\beta$.
We will find the latter will be fixed by choice of $m_0$ if we require Higgsino DM, and the former will be fixed by the anomaly-mediated formula for gaugino mass for wino DM.  We will explore the implications for relaxing the assumption that the soft masses of the Higgs fields take on this universal value. We will also explore the implications of allowing the relationship between the scale for anomaly mediation and $m_0$ to vary.\footnote{Namely, we add an enhancement factor to the gaugino masses and trilinear couplings.}  Lastly, the physically relevant CP-violating phase will be assumed to be order one, though smaller values will be considered in the case of Higgsino DM.

Similar split SUSY models have been studied previously~\cite{wells_implications_2003,Pierce:2004mk,arkani-hamed_well-tempered_2006, Baer:2016ucr, kowalska_discreet_2018}, some focusing on Higgsino DM~\cite{Vereshkov:2005ad,Beylin:2007yt,Baer:2011ec,Co:2021ion} or wino DM~\cite{wells_pev-scale_2005,Cohen_wino_2013,Fan:2013faa}.  We identify regions in parameter space that reproduce the observed Higgs boson mass.
We will discuss in Section~\ref{sec:EW} how in the Higgsino DM case, EW symmetry breaking requirements limit our model into a region with a somewhat tuned,   unified scalar mass, and we will analyze the general implications of the SM Higgs boson mass, see also Ref.~\cite{Arvanitaki:2012ps}. In Section~\ref{sec:higgsino} we will discuss how electric dipole moment (EDM) and direct detection experiments may reach the entirety of this space for Higgsino DM. In Section~\ref{sec:wino} we will similarly discuss how EDM experiments may reach the entire space of wino DM models. All of this work will be in the service of our central point, which is to elaborate how this motivated model of supersymmetry will either be discovered or entirely ruled out in the near future.

\section{Fine tuning in electroweak symmetry breaking for light Higgsinos}
\label{sec:EW}

\subsection{Constraint from electroweak symmetry breaking}
\label{sec:constraints}

EW symmetry breaking imposes relationships between the Higgs boson soft mass parameters  $m_{H_u},m_{H_d},$ the $Z$ boson mass, the Higgsino mass parameter $\mu$, and  $B\mu$. In our setup, the former two of these are set at high scales to the unified scalar scale, $m_0$. As we will see, to achieve the observed $Z$-boson mass, $\mu$ and $B$ are comparable to $m_{0}$ except in particular circumstances.   We are interested in the possibility that $\mu \ll m_{0}$ so as to reproduce Higgsino dark matter. We will discuss the precise relationship this requires between $m_0$ and $\tan\beta$ (see also Ref.~\cite{Arvanitaki:2012ps}), along with the constraints imposed by the SM Higgs mass, and how these requirements can be eased by relaxing the requirement of unification of the scalar sfermion masses and the SUSY-breaking Higgs boson masses.

\begin{figure}[t]
    \centering
    \includegraphics[width=\linewidth]{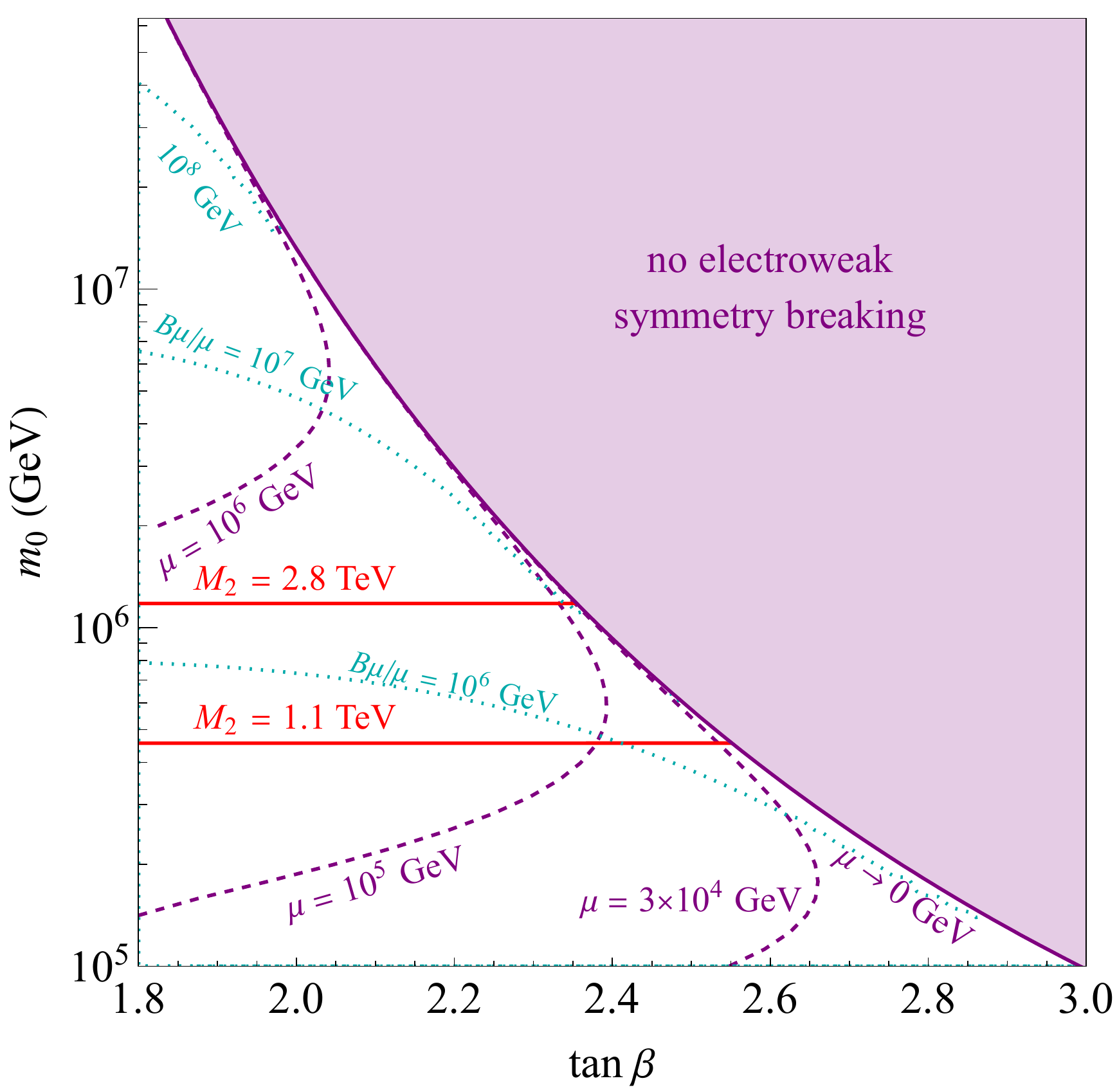}
    \caption{Contours of $\mu$ (purple), $M_2$ (red), and $B\mu/\mu$ (cyan), where the gaugino masses are set by AMSB at high scale and $\mu$ is set as needed for EW symmetry breaking, assuming $m_{H_u}$ and $m_{H_d}$ are $m_0$ at the GUT scale. EW symmetry is unbroken in the purple shaded region. The contour with $\mu = 1.1$ TeV, of interest for dark matter, is indistinguishable from the $\mu \rightarrow 0$ boundary, see text for details.}
    \label{fig:mu_m2_contours}
\end{figure}

The potential for the neutral Higgs bosons is
\begin{equation}
\begin{split}
    V_{\text{MSSM}}=&\pr{|\mu|^2 + m_{H_u}^2}\left|H_u^0 \right|^2+\pr{|\mu|^2 + m_{H_d}^2}\left|H_d^0 \right|^2 \\ 
    &- \pr{B\mu H_u^0H^0_d + \text{c.c.}} \\
    &+\frac{1}{8}\pr{g^2+g'^2}\pr{\left|H_u^0\right|^2-\left|H_d^0 \right|^2}^2.
\end{split}
    \label{eq:higgsPotential}
\end{equation}
The minimum of this potential sets the two Higgs vacuum expectation values (vevs), with the Standard Model vev $v= 246$ GeV given by $v^2 = v_u^2+v_d^2$, and the ratio between them defining $\beta$ such that $\tan\beta \equiv v_u/v_d$. One can find the minimum for this potential at the respective Higgs vevs, giving
\begin{equation}
    \begin{aligned}
&m_{H_{u}}^{2}+|\mu|^{2}-B\mu \cot \beta-\frac{m_Z^{2}}{2} \cos{2 \beta}=0, \\
&m_{H_{d}}^{2}+|\mu|^{2}-B\mu \tan \beta+\frac{m_Z^{2}}{2} \cos{2 \beta}=0.
\label{eq:EW_SUSY_Calc}
\end{aligned}
\end{equation}
These equations can  be used to solve for $\mu$ and $B\mu$ in terms of $M_{Z}$, $\tan \beta$  and quantities set by our chosen unified scalar mass:
\begin{equation}
   |\mu|^{2} =\frac{1}{2}\pr{ \frac{\left|m_{H_{d}}^{2}-m_{H_{u}}^{2}\right|}{\sqrt{1-\sin ^{2}{2 \beta}}}-m_{H_{u}}^{2}-m_{H_{d}}^{2}-m_Z^2},
   \label{eq:mu_calc}
\end{equation}
\begin{equation}
       B \mu =\pr{\frac{m_{H_{u}}^{2}}{2}+\frac{m_{H_{d}}^{2}}{2}+|\mu|^{2}}\sin{2 \beta}.
   \label{eq:b_calc}
\end{equation}
Reproducing the observed Higgs boson mass requires stop masses (and hence $m_{0}$) 
 much larger than the EW scale.  We therefore expect $m_{H_{u}}$ and $m_{H_{d}}$ to be similarly large, and  Eq.~(\ref{eq:mu_calc}) then indicates that $\mu$ will generally be on the order of the unified scalar mass, $m_0$, unless cancellations occur.
In Fig.~\ref{fig:mu_m2_contours}, we show contours of $\mu$ in the $\tan \beta$-$m_0$ plane.  Indeed, over much of the plane $\mu$ is of the same order as $m_{0}$.  
However, up to corrections of the order of $m_Z$, when $m_{H_u}$ runs to a value at the symmetry breaking scale given by
\begin{equation}
    m_{H_u} (m_{\rm SUSY}) = \frac{m_{H_d} (m_{\rm SUSY})}{\tan\beta} \simeq \frac{m_0}{\tan\beta},
    \label{eq:tuning}
\end{equation}
$\mu$ vanishes at $m_{\rm SUSY}$.
Here $m_{\rm SUSY}$ indicates the parameters are run from their GUT-scale value $m_0$ down to the SUSY scale,
\begin{equation}
    m_{\rm SUSY} = \sqrt{m_{\Tilde{t}_L}m_{\Tilde{t}_R}}.
\end{equation}
Since $m_{H_d}$ runs negligibly, it is approximately $m_0$ at the SUSY scale. For $m_{H_u}$ greater than the condition in Eq.~(\ref{eq:tuning}), EW symmetry is preserved due to the large values of the SUSY breaking Higgs masses. For masses below this condition, EW symmetry is broken either by the $B\mu$ term or by $m_{H_u}^2$ running negative. Within the purple region of Fig.~\ref{fig:mu_m2_contours}, $m_{H_u}^2$ runs sufficiently slowly relative to the value of $m_0$ that EW symmetry is in this way preserved. At the boundary, the contributions from $m_{H_u}^2$ and $m_{H_d}^2$ in Eq.~(\ref{eq:mu_calc}) cancel, leaving very small $\mu$. For Higgsino-like thermal DM, the Higgsino mass is $ 1.1 \pm 0.1 \TeV \ll m_0,$ in a split SUSY model~\cite{Profumo_Statistical_2004,giudice_split_2005,Pierce:2004mk,Hryczuk:2010zi}. 

Shifts in $m_0$ at the GUT scale tend to cause similarly sized shifts in the value of $m_{H_u}(m_{SUSY})$.  Because we are interested in $m_{0} \gg \mu$, the allowed fractional change in $m_{0}$ that maintains a small $\mu$ is quite small for fixed $ \tan \beta$. This is why the curves for $\mu = 0$ and for the desired DM mass $\mu = 1.1$ TeV are imperceptibly close to one another in Fig.~\ref{fig:mu_m2_contours}.  In this figure, we also show two contours for $M_{2}.$ The contour for $M_{2} = 2.8 \TeV$, determined by the AMSB relation from $m_{0} \simeq m_{3/2}$ should be taken as indicative of the parameter space that reproduces wino DM~\cite{Hisano:2006nn,Cirelli:2007xd,Hryczuk:2010zi}.  If the dark matter is to be Higgsino with mass 1.1 TeV, the wino must have a larger mass, so $m_{0}$ must lie above the $M_{2} = 1.1$ TeV contour.

\begin{figure}[t]
    \centering
    \includegraphics[width=0.9\columnwidth]{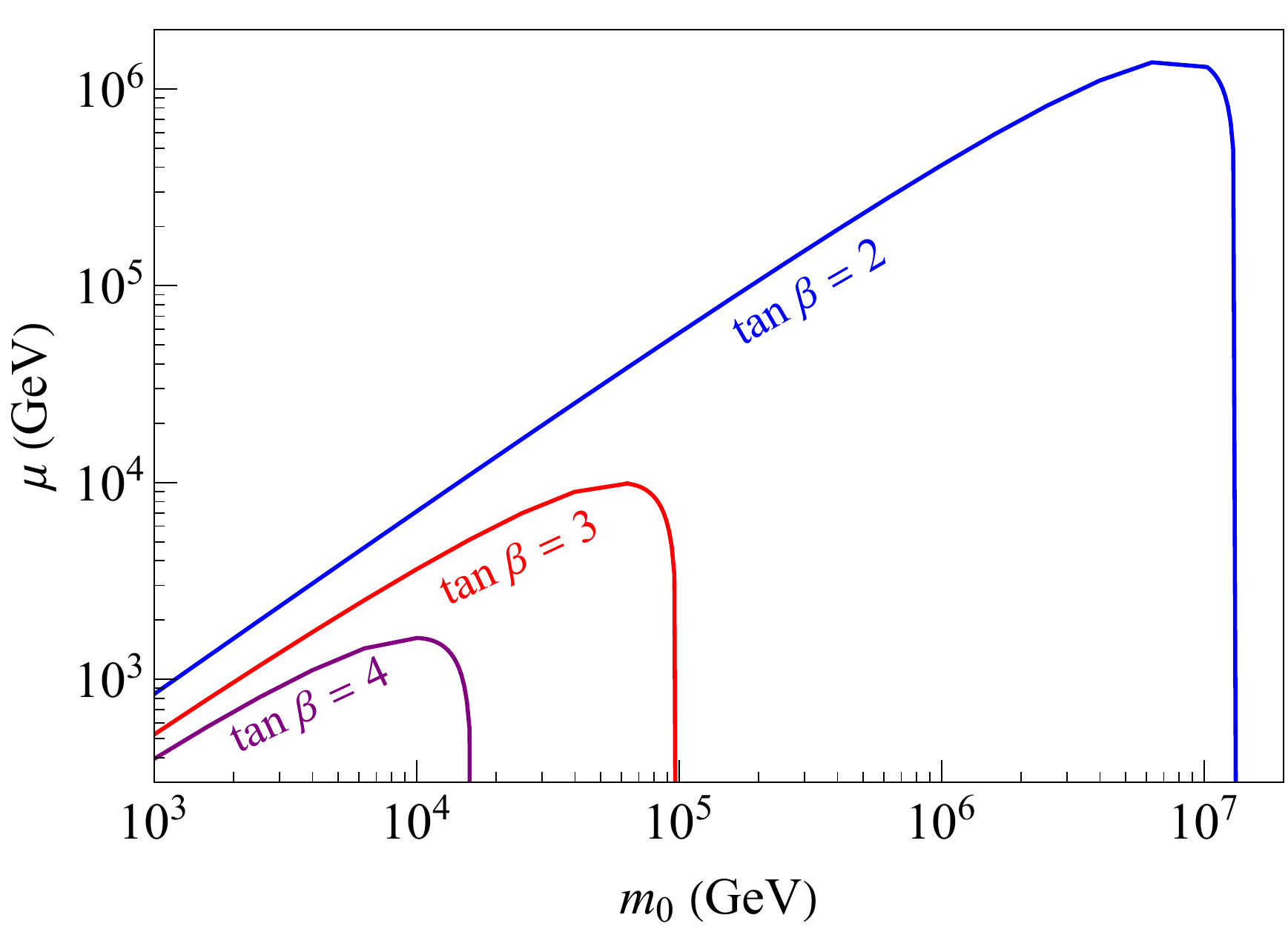}
    \includegraphics[width=0.9\columnwidth]{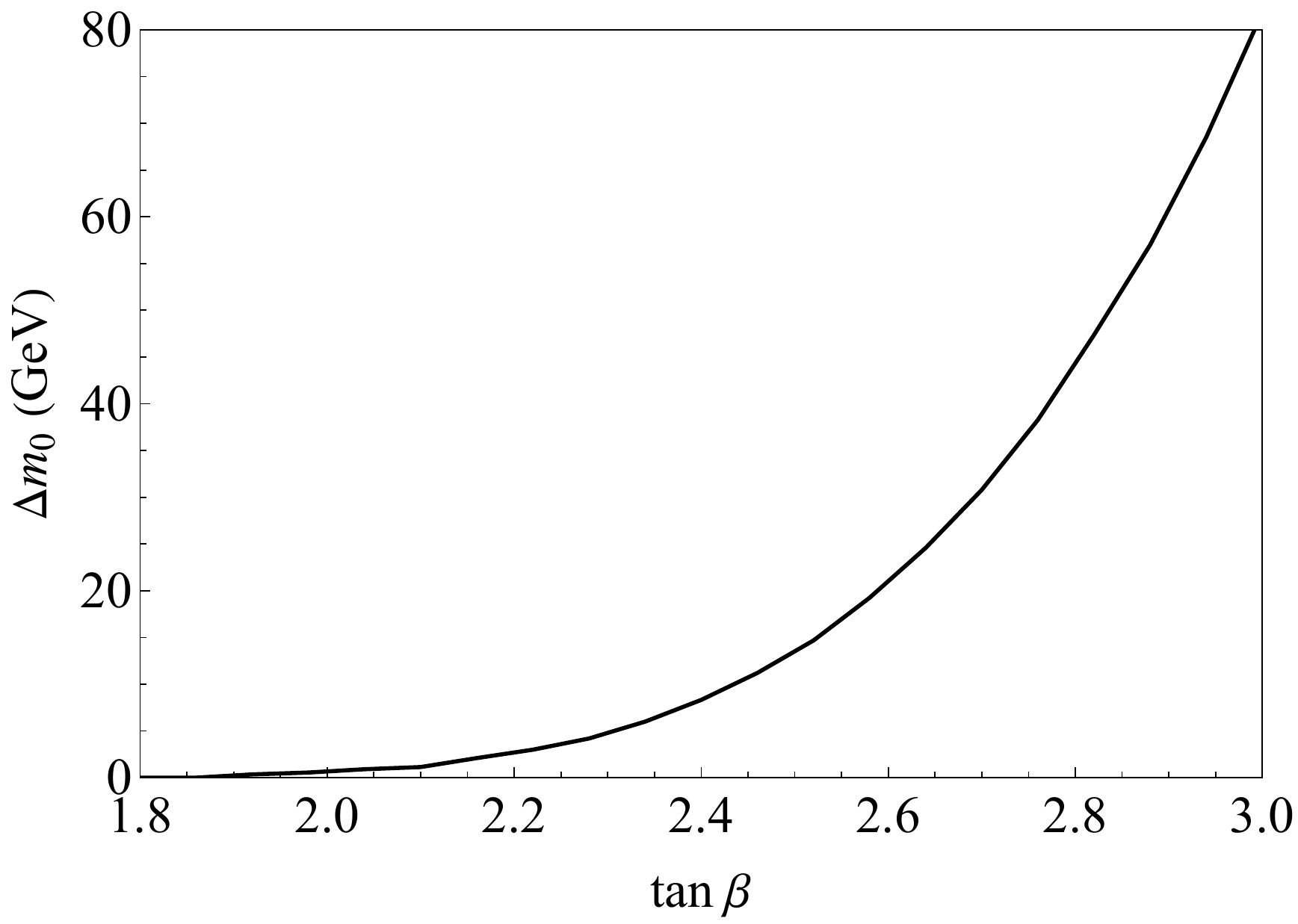}
    \caption{(Upper) the value for $\mu$ needed for EW symmetry breaking given $m_0$ and $\tan \beta$. The curve rises following roughly $\mu^2 \approx m_0^2/\pr{\tan^2\beta-1}$ then drops off rapidly at large $m_0$, as $m_{H_u}$ runs close to $m_{H_d}/\tan\beta$ at the SUSY scale. At any larger $m_0$,
    EW symmetry breaking fails. (Lower) $\Delta m_0$ as the width of $m_0$ that corresponds to 1 TeV $\le \mu \le $ 1.2 TeV. This narrow parameter space is a result of requiring an appropriate $\mu$ for Higgsino DM. As $\tan\beta$ gets larger, the range of possible $m_0$ values for small $\mu$ gets thinner.}
    \label{fig:muOverM0}
\end{figure}

As alluded to above, in this setup, Higgsino dark matter requires a particular value of $m_{0}$ for each value of $\tan \beta$. This is emphasized in Fig.~\ref{fig:muOverM0}, where we have displayed the value of $\mu$ as a function of $m_{0}$ as determined by the relationship of Eq.~(\ref{eq:mu_calc}) for different choices of $\tan \beta$.
Note that $\mu$ falls off rapidly as the tuning condition is achieved. We can rewrite Eq.~(\ref{eq:mu_calc}) as
\begin{equation}
    m_{H_d} - m_{H_u} \tan\beta \approx \frac{|\mu|^2 \pr{\tan^2\beta - 1}}{m_{H_d}+ m_{H_u}\tan\beta} \sim \frac{|\mu|^2}{m_0}.
    \label{eq:mu_mHu_narrowness}
\end{equation}
This implies that as $\tan\beta$ decreases (and hence the $m_0$ value needed for small $\mu$ gets larger), the difference of $ m_{H_d}$ and $m_{H_u}\tan\beta $ is enhanced by a factor of $m_0/\mu$, so the degree to which they must match to maintain small $\mu$ gets larger. This is reflected in the lower panel in Fig.~\ref{fig:muOverM0}, as the range of values for $m_0$ that reproduces a Higgsino mass within 10\% of the value needed for DM at a given $\tan\beta$ value has a width that grows narrower at smaller $\tan\beta$.

The value for $B$ within the small $\mu$ regime can be intuited by simplifying Eq.~(\ref{eq:b_calc}). For $\mu \ll m_0$, $m_{H_d} \simeq m_0$ and $m_{H_u} \simeq m_0/\tan\beta$, we get
\begin{equation}
    B \sim \frac{m_0^2}{\mu} \cot{\beta}.
\end{equation}
Indeed, even far from the EW symmetry preserving region, $B$ follows roughly as $m_0$ when $\mu$ gets large, but near that region, $\mu$ falls off. In particular, along the contour of $\mu = 1.1 \TeV$, imperceptibly close to the $\mu=0$ contour in Fig.~\ref{fig:mu_m2_contours}, $B$ ranges from $10^7$ GeV to $10^{11}$ GeV along the regime of interest. This corresponds to ($10^2$-$10^4$)$\times m_0$, which is a relatively conspicuous enhancement.

\subsection{Constraint from Higgs boson mass}

\begin{figure}[t]
    \centering
    \includegraphics[height=\linewidth]{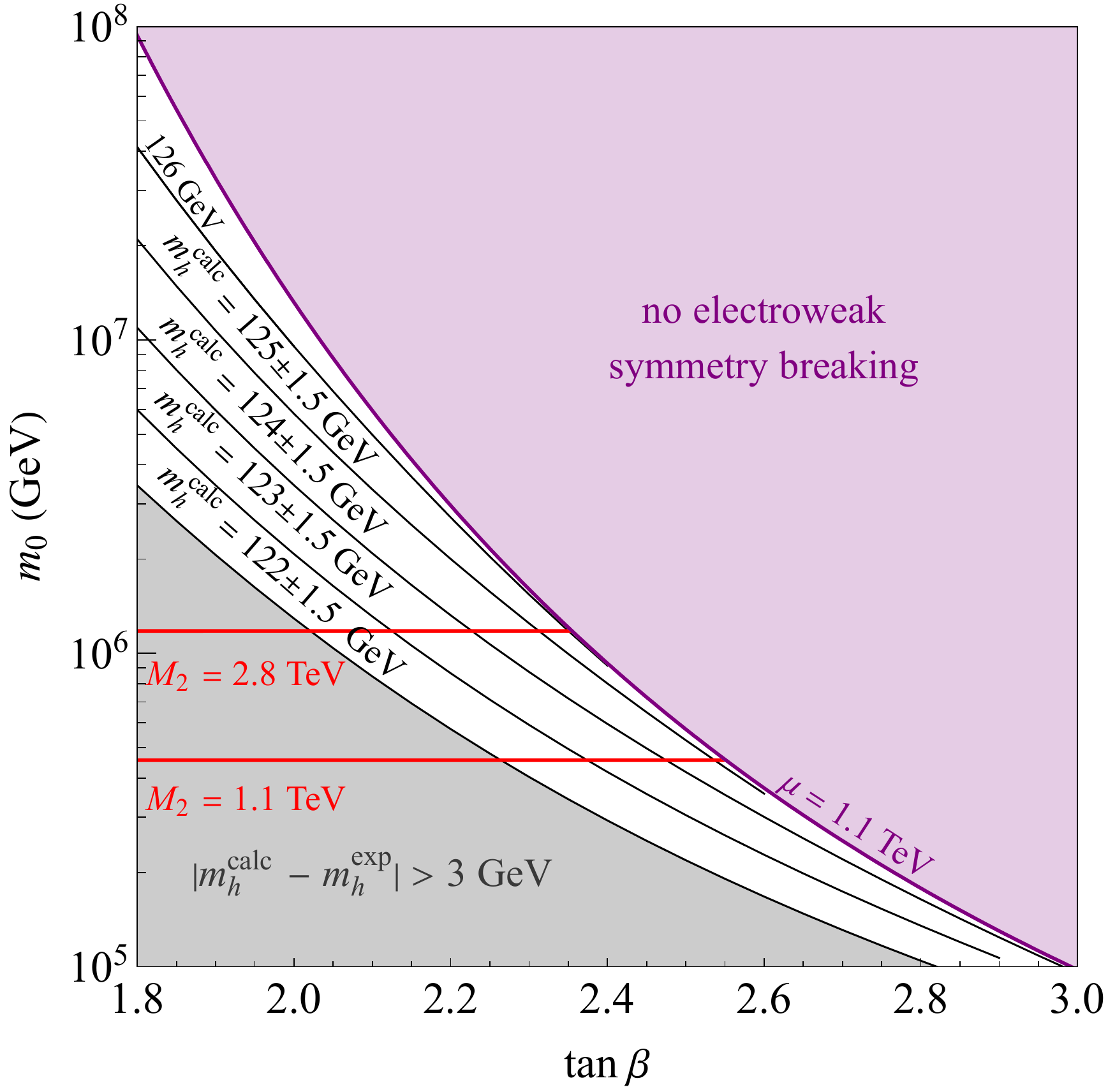}
    \caption{The parameter space of interest for split SUSY with AMSB gauginos, where the black contours are labeled (text above each curve) by the computed mass of the SM Higgs boson, and the gray region is then excluded at 2$\sigma$. The red lines are contours of constant $M_2$, the purple contour is constant $\mu=1.1$ TeV, with a thin, not discernible separation to the boundary of the purple region, in which EW symmetry is not broken, making the theory no longer viable.}
    \label{fig:higgs_limits}
\end{figure}

In addition to discussing constraints from EW symmetry breaking, we must also consider implications from the constraint of obtaining the 125 GeV mass of the SM-like Higgs boson. In the MSSM, this mass can be calculated in terms of known SM quantities and 
SUSY loop corrections. We compute the Higgs mass and theoretical errors using SusyHD~\cite{PardoVega:2015eno}, a 
package designed to calculate the Higgs mass to two-loop order within an MSSM model with parameters specified at the SUSY scale.  We take the SUSY breaking scalar masses to be $m_0$ at the GUT scale and take gaugino masses set by the AMSB relations with $m_{0} = m_{3/2}$, and find the required SUSY scale  masses through renormalization group evolution. We find  $\tan\beta$, $\mu$, and the pseudoscalar mass $m_{A_0}$ through EW symmetry breaking requirements, and calculate the unified trilinear coupling described by anomaly mediation~\cite{randall_out_1999}.   

Given the resulting SUSY scale masses, SusyHD finds the mass of the SM-like Higgs boson and gives the remaining theoretical uncertainty to the Higgs mass calculation, which is close to 1 GeV across our parameter space of interest. Additional parametric sources of uncertainty, dominantly the experimental uncertainty in the mass of the top quark, $172.76 \pm 0.30$ GeV~\cite{ParticleDataGroup:2020ssz}, 
contribute an additional uncertainty of approximately 1 GeV. Accounting for these sources of uncertainty, we restrict our model to regions with a calculated Higgs mass of $125 \pm  3$ GeV at $2\sigma$, shown in contours over the full scope of the parameter space of $m_0$ and $\tan\beta$ in Fig.~\ref{fig:higgs_limits}. For the Higgsino DM case (along the boundary of the purple region), the viable parameter space is constrained to the segment above the $M_2 = 1.1 \TeV$ contour so the Higgsino is in fact the LSP. Values of $m_0$ higher than $10^8 \GeV$ that have $\tan\beta < 1.8$ are less motivated due to concerns over the perturbativity of the top quark Yukawa coupling and the degree of gauge coupling unification.
For wino DM (long the $M_2 = 2.8 \TeV$ contour), the Higgs mass constraints also significantly limit the available region to $\tan\beta \gtrsim 2$. On the other hand, the requirement for a wino LSP requires $\tan\beta \lesssim 2.35$.

\subsection{Easing the tuning with Higgs sector freedom}
In Sec.~\ref{sec:constraints}, we found that our ansatz of universal scalar masses, combined with the requirement of Higgsino DM, led to both very particular choices of $m_0$ and large values for $B/m_0$. 
Here we explore the robustness of this conclusion when the Higgs sector parameters $m_{H_u}$ and $m_{H_d}$ vary independently of the universal scalar mass.

These parameters cannot take on arbitrary values.   
In particular, $m_{H_{u}}^2$ cannot be too much larger than $m_0^2$; otherwise it will induce tachyonic masses for the stop squarks.  To see this,  note that at large the  $m_{H_{u}}^2$, the one-loop beta function of the right-handed stop squark soft mass is dominated by the up-type Higgs mass,
\begin{equation}
    \beta_{m_{u}^{2}}^{(1)}=4 m_{H_{u}}^{2} |\lambda_t|^2 +  \cdots,
\end{equation}
for top Yukawa coupling $\lambda_t$. This running implies any increase in $m_{H_{u}}^2$ will cause $m_u^2$ to deviate farther from $m_0^2$ due to running from the GUT scale to the SUSY scale.
For unified sfermion mass values around the PeV scale specified at a GUT scale $\sim 2 \times 10^{16}$ GeV, $m_{H_{u}}^{2}\gtrsim 3m_0^2$ will imply tachyonic right-handed stop squarks at the SUSY scale. There is a similar effect with a heavy down-type Higgs, but the running is suppressed by $g_1^2/ (5 |\lambda_t|^2)$ as compared to the up-type Higgs effect. 

The Higgs soft masses also cannot be made arbitrarily small. If $m_{H_{u}}^2$ runs below zero at the SUSY scale, then Eq.~(\ref{eq:mu_calc}) forbids $\mu$ parametrically smaller than $\left| m_{H_u} (m_{\rm SUSY})\right|\sim m_0$. 
This observation places a lower bound on the GUT scale value of $m_{H_{u}}$ that will be roughly 90\% of $m_0$. To aid in intuition, we observe that $m_{H_{u}}^2$ runs roughly linearly with the log of the energy scale between the SUSY scale and the GUT scale for $m_0$ between $10^5$ and $10^8$ GeV.\footnote{While this holds true beyond this range, as seen in the previous sections this is the range between models that are ruled out by colliders and models that require $\tan\beta$ small enough to imply perturbativity issues in the top Yukawa} 
It runs through $m_0^2/\tan^2\beta \approx m_0^2/4$ near the SUSY scale. This means for it to pass through zero at the SUSY scale, $m_{H_{u}}^2$ would have to be reduced by roughly one quarter its value, so $m_{H_{u}} \approx 0.87 m_0$. While this description ignores   details, it is accurate within roughly 10\%. For any GUT scale value of $m_{H_{u}}^2$ below this, $m_{H_{u}}^2$ would run below zero at the SUSY scale, which would forbid a Higgsino mass significantly lighter than $m_0$ as is required for Higgsino DM.

\begin{figure}[t]
    \centering
    \includegraphics[width=0.875\linewidth]{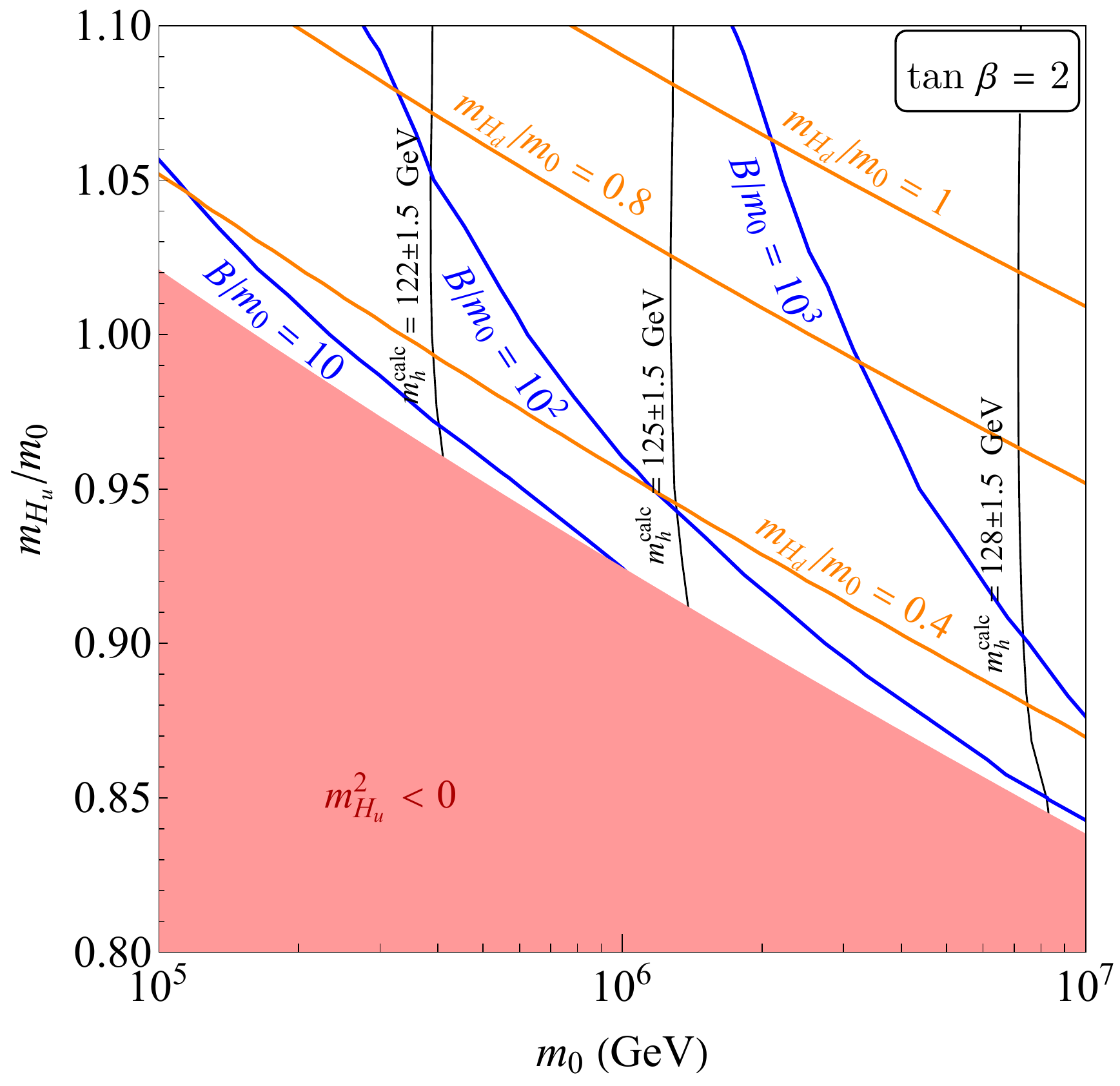}
    \includegraphics[width=0.875\linewidth]{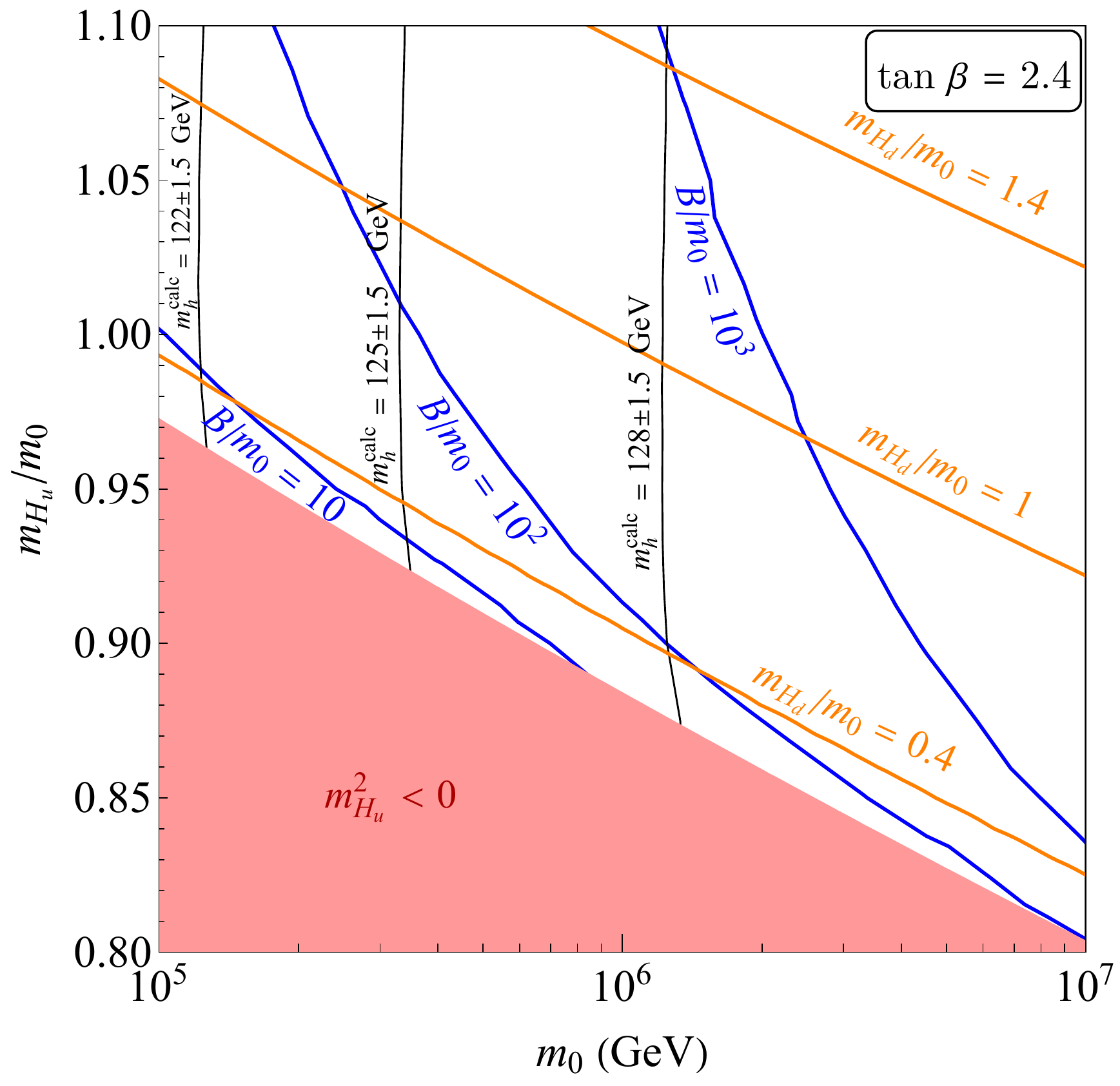}
    \caption{Contours of $B/m_0$ with $\tan\beta = $ 2 (upper) or 2.4 (lower). At each value of $m_0$ and $m_{H_u}$, $m_{H_d}$ is chosen to give $\mu = 1.1$ TeV. All three of these input parameters are taken at the GUT scale, $2 \times 10^{16}$ GeV and run down to find $\mu$ and $B$ at the SUSY scale based on the requirements for EW symmetry breaking. $m_{H_u}^2 < 0$ is excluded as $\mu$ is near $m_0^2$ in this regime, and the region significantly above the displayed range is excluded due to tachyonic stop squarks. Contours of constant $m_{H_d}/m_0$ are included in orange to aid in intuition, and the outer bounds and preferred value for the calculated SM Higgs mass are included as well. For larger $\tan\beta$, both the lower and upper bounds for $m_{H_u}/m_0$ shift further down, and the calculated SM Higgs mass increases, pushing the allowed region left.}
    \label{fig:floatinghiggs}
\end{figure}

Allowing $m_{H_{u}}^2$ and $m_{H_{d}}^2$ to vary while imposing the DM Higgsino constraint $\mu = 1.1$ TeV, we arrive at the contours shown in Fig.~\ref{fig:floatinghiggs}. Taking advantage of the freedom in the Higgs soft masses,  $m_0$ need no longer be specified as precisely, but as can be seen from the figure, the requirements on $m_{H_d}$ at a given $m_{H_u}, m_0$, and $\tan\beta$ remain strict. 
The value for $B$ varies substantially as the SUSY scale Higgs sector parameters vary. In particular, there is a narrow regime in which $B$ is of similar order to $m_0$, to the left and below the leftmost blue contours in Fig.~\ref{fig:floatinghiggs}. 
By comparing the two panels, we can see, as $\tan\beta$ gets larger, the region allowed by the Higgs boson mass constraints (between the black contours)  is at lower $m_0$ and smaller $B/m_0$ value.

\section{Experimental searches for Higgsino dark matter}
\label{sec:higgsino}

We will focus on two experimental probes for split SUSY with AMSB gauginos and thermal Higgsino DM models, the direct detection of dark matter and searches for the electron EDM.  We will discuss how next generation experiments have the potential to explore the entirety of the available parameter space, which itself is constrained by requiring a 125 GeV Higgs boson.

The interaction between DM and nucleons on which direct detection depends is dominated by the tree-level scattering mediated by the the light Higgs boson for Higgsino and wino DM~\cite{jungman_supersymmetric_1996, hisano_direct_2011, hisano_direct_2013}.  We calculated the nucleon scattering cross section with \micromegas\cite{Belanger:2010pz} and compared against limits placed by Xenon1T~\cite{aprile_dark_2018} and against the expected reach for LUX-ZEPLIN (LZ)~\cite{LZ_projected_2020}. Loop order effects tend to be subdominant in this regime~\cite{Chen:2019gtm,hill_wimp-nucleon_2014}.  
For $m_0 \gtrsim 10^6$ GeV, we have $M_2 \gg \mu \gg m_Z$, which implies a small wino-Higgsino mixing angle, so the tree-level dark matter-nucleon scattering cross section is highly suppressed.  
Interactions mediated by \textit{Z} bosons are also suppressed by a similar factor of $ m_Z / \mu$. While they have a higher dark matter-nucleon cross section, 
the spin independent interactions result in stronger limits~\cite{Belanger_Discriminating_2009,Cohen_Correlation_2010,cheung_prospects_2013} because the Higgs-mediated interactions described here scale up with the size of the atomic nucleus due to coherent interactions across nucleons.

For the electron EDM, the dominant contribution at this level is from two-loop, Barr-Zee diagrams~\cite{Barr_zee_edm}, in which a chargino runs in an internal loop, connected by EW bosons to a lepton line. One-loop diagrams have a much smaller contribution for scalar masses above 10 TeV~\cite{delAguila:1983dfr, cesarotti_interpreting_2019}. The degree of CP violation in such 
two-loop contributions can be simplified to two irreducible phases,  $\operatorname{Im}\left(\tilde{g}_{u}^{\prime *} \tilde{g}_{d}^{\prime *} M_{1} \mu\right)$ and $\operatorname{Im}\left(\tilde{g}_{u}^{*} \tilde{g}_{d}^{*} M_{2} \mu\right)$, following the notation of Ref.~\cite{arkani-hamed_aspects_2005}. These will be referred to as $\phi_1$ and $\phi_2$ respectively, as they can be thought of as phases on the respective gaugino masses. In this work, the  gaugino masses are proportional to the gravitino mass, so both of these phases are the same. 
Specific calculations of these diagrams can be found in Ref.~\cite{giudice_electric_2006}. The electron EDM is compared against the limit placed by ACME II~\cite{ACME_improved_2018} and against the expected reach of Advanced ACME~\cite{wu_metastable_2020,panda_attaining_2019}, at roughly one order of magnitude lower EDM values than ACME II.

For Higgsino-like DM, $\mu \simeq 1.1$ TeV, which requires a precise relation between $m_0$ and $\tan\beta$ as shown in Fig.~\ref{fig:mu_m2_contours}.
The range of $m_0$ values for a given $\tan\beta$ with $\mu$ set to allow thermal Higgsino DM has width $\Delta m_0 \lesssim\mathcal{O}\pr{100}$ GeV as seen in the lower panel of Fig.~\ref{fig:muOverM0}. The fractional width of this range of $m_0$ values is at most $0.1\%$. Similarly, the precision of the requirement that Higgsino DM thermally produce the relic abundance is an $\mathcal{O}\pr{10\%}$ requirement on $\mu$. Neither of these ranges for $m_0$ or $\mu$ offers substantial changes to our experimental limits.

\begin{figure}[t]
    \centering
    \includegraphics[width=\linewidth]{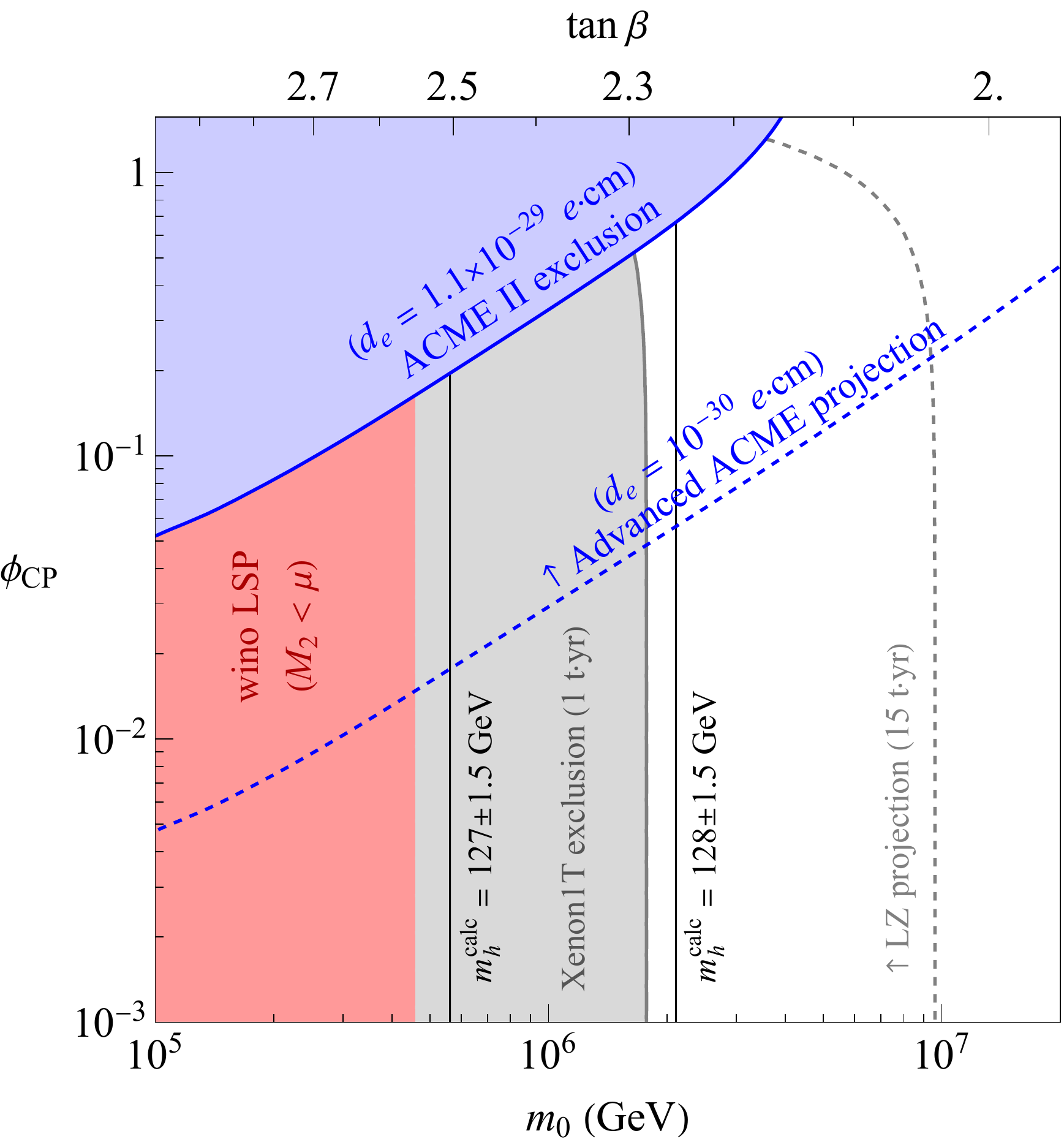}
    \caption{Constraints for models with thermally produced Higgsinos composing the entirety of DM, with constraints marked as colored solid lines and filled regions denoting exclusions, and dashed lines denoting projections. $d_e$ represents the electron EDM, Xenon1T (LZ) is the direct detection bound (prospect), $m_h^{\rm calc}$ is the calculated Higgs boson mass, and the wino LSP region is shaded as our assumption of Higgsino DM is false. As the regime for $m_0$ that gives the needed values for $\mu$ for a given $\tan\beta$ is a narrow, monotonic band, we display the corresponding $\tan\beta$ value as the top $x$-axis. }
    \label{fig:DM_mu_constraints}
\end{figure}

One of the advantages of split supersymmetry scenarios is that flavor and SUSY CP problems are relaxed, in which case one might suspect that the CP violating phase, $\phi_{\rm CP} = \phi_1 = \phi_2$, is expected to be of order one, though for this study we include a range down to $10^{-3}$.  The full set of constraints and discovery potential from existing and planned experiments within this space is shown in Fig.~\ref{fig:DM_mu_constraints}. The calculated mass of the Higgs boson is relatively large but is within a 95\% confidence interval from theoretical errors and uncertainties propagated from the top quark mass. The electron EDM is sufficiently high that, for maximal CP violation, the entirety of the parameter space is excluded due to limits from  ACME II~\cite{ACME_improved_2018}. The projections and limits for electron EDM experiments weaken as $\phi_{\rm CP}$ decreases since the electron EDM scales with $\sin\phi_{\rm CP}$. Advanced ACME will reach into $\mathcal{O}(0.1)$ CP violation. 

The dark matter direct detection projections and limits, on the other hand, scale as $\cos\phi_{\rm CP}$ and strengthen to an asymptotic maximal reach along $m_0$ at small $\phi_{\rm CP}$, as we can see in Fig.~\ref{fig:DM_mu_constraints}, for $\phi_{\mathrm CP} \lesssim 0.1$, Xenon1T exclusion and LZ projections reach $1.7 \times 10^6$ GeV and $1.0\times 10^7$ GeV respectively. This builds an interesting complementarity, as further refinement on the measured top quark mass, which would refine the Higgs mass limits, could reduce the error on the calculated Higgs mass as far as to $\approx 1$ GeV, which would reduce the 95\% exclusion curve in Fig.~\ref{fig:DM_mu_constraints} to 127 GeV, well inside the Xenon1T and ACME II exclusion regions. Alternatively, results from LZ and Advanced ACME would have the potential to reach the entirety of the remaining parameter space, with discovery potential well beyond the 95\% exclusion region for calculated Higgs mass. In either case, we expect near future experiments to provide a definite answer as to the validity of this model.

In the case with non-universal Higgs scalar masses, there is no longer as tight a mapping between the scalar masses and $\tan \beta$ for the Higgsino dark matter case.  However, Fig.~\ref{fig:DM_mu_constraints} still roughly applies, using now only the $m_{0}$ as the $x$-axis.  Throughout the range of $m_{0}$ shown, $\tan \beta$ varies over only a modest range, so the regions of dark matter direct detection and  EDM constraints are still approximately valid.

\section{Experimental searches for wino dark matter}
\label{sec:wino}

\begin{figure}[t]
    \centering
    \includegraphics[width=\linewidth]{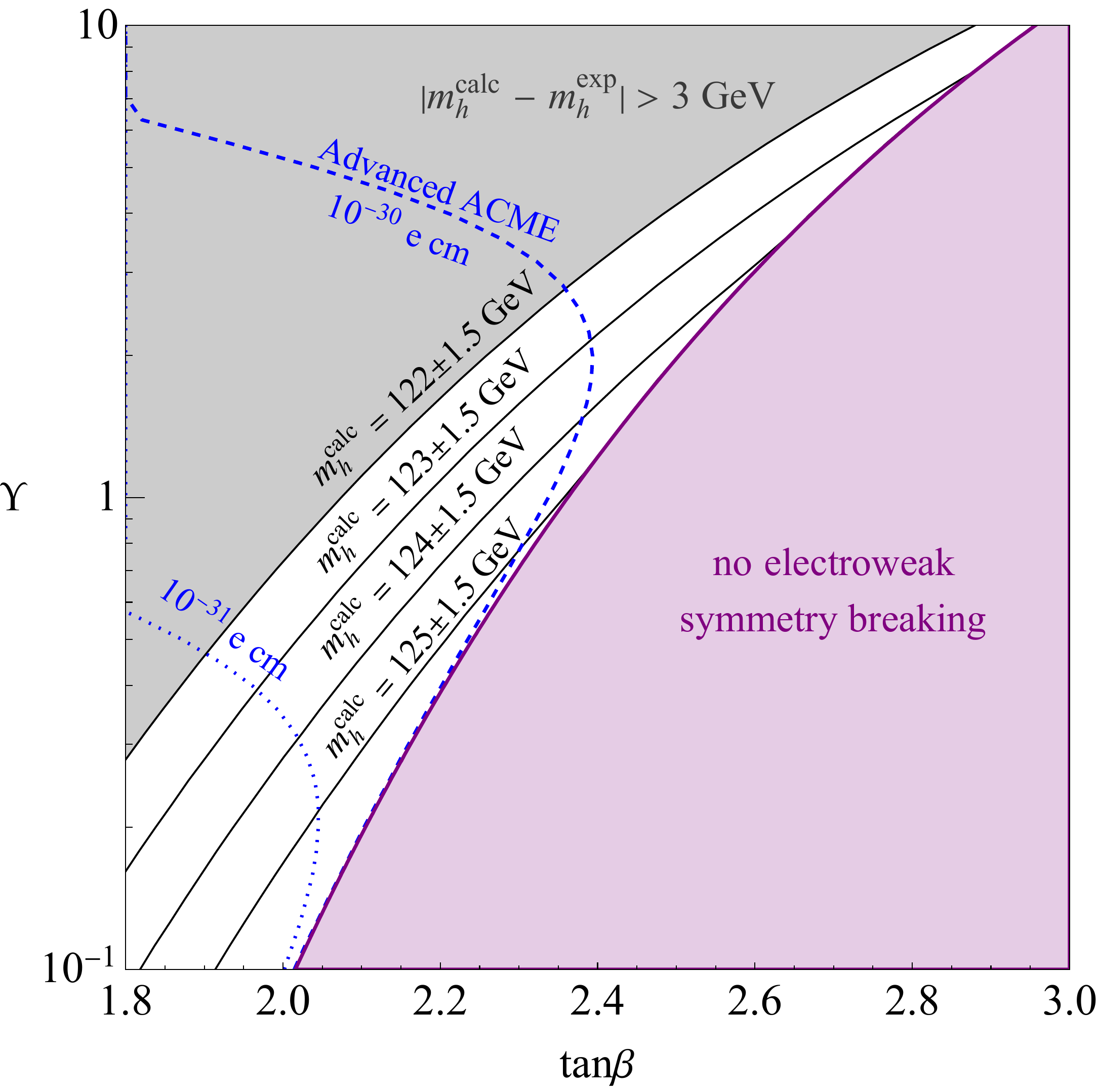}
    \caption{Constraints for models with thermally produced winos composing the entirety of DM. The gray region is excluded by Higgs mass measurements, and the purple region is excluded as EW symmetry is preserved. The dashed curve is the projected reach for a future  electron EDM experiment, the dotted curve below is a contour of electron EDM attainable by more distant, unspecified future experiments, and the $m_h^{\rm calc}$ curves are contours of calculated Higgs mass. Electron EDMs are calculated assuming maximal CP violation.}
    \label{fig:DM_M2_constraints}
\end{figure}

In the case of wino DM, the limits from next generation electron EDM experiments are more modest, but are not far from the sensitivity needed to reach the entire space. 
By setting  $M_2 = 2.8 \TeV$ to thermally produce the right DM relic density~\cite{Hisano:2006nn,Cirelli:2007xd,Hryczuk:2010zi}, up to $\mathcal{O}\pr{10\%}$ precision, we require $m_0$ to be 1.2 PeV. This level of imprecision does not change any of our qualitative conclusions appreciably. As we can see in Fig.~\ref{fig:higgs_limits}, requiring a $\sim 125$ GeV Higgs boson and EW symmetry breaking places the constraints $\tan\beta > 2.02$ and $\tan\beta < 2.35$ respectively. The direct detection limits and projections are negligible here, as we will discuss later in this section, and Advanced ACME projections will reach $\tan\beta > 2.33$, but the entire regime allowed by the Higgs mass constraints will have electron EDM $> 10^{-31} \sin\phi_{\rm CP}$  $e$ cm, so will be discoverable in next-to-next generation experiments.

We can make a more interesting exploration by loosening the rigidity of our model. 
In particular, we allow our scales to float relative to one another in the following way. Given the gravitino mass, we set the scale used for AMSB masses, relevant for gauginos, and the scale for the scalar masses as
\begin{equation}
    m_{\rm AMSB} =\chi_1m_{3/2}, \quad   \quad  
    m_0 = \chi_2 m_{3/2},
\end{equation}
for some suppression or enhancement factors $\chi_i$. We can then define the ratio
\begin{equation}
    \Upsilon = \frac{\chi_1}{\chi_2},
\end{equation}
as a net relative enhancement factor of the gaugino masses with respect to $m_0$, which contains all measurable effects from $\chi_1$ and $\chi_2$. In the previous sections, the scale for anomaly-mediated mass contributions is taken the same as $m_0$, i.e., $\Upsilon = 1$. In practical terms, $\Upsilon$ will vary inversely with $m_0$ for fixed wino mass, so we can explore the effects of different scalar mass scales, and thereby different Higgsino masses, on the wino-like DM regime. The effects of different $\Upsilon$ are shown in Fig.~\ref{fig:DM_M2_constraints}.

For large values of $\Upsilon$, Advanced ACME stands to reach the entire available parameter space after Higgs mass constraints. To reach the entirety of the space will require more futuristic EDM experiments with 1-2 orders of magnitude more precision.

In contrast to the Higgsino DM case, most of the parameter space in Fig.~\ref{fig:DM_M2_constraints} is difficult to reach via direct detection. This is because the cross section falls off rapidly with the mixing angle between gauginos and Higgsinos, as over most of this figure, $\mu\sim m_0 \sim 10^3\times  M_2$. For $M_2 = 2.8$ TeV, Xenon1T only constrains models with $\mu \lesssim 3$ TeV, and LZ reach extends this to $\mu \lesssim 5$ TeV. The direct detection limits and projections are then in a narrow band along the boundary of the purple region of Fig.~\ref{fig:DM_M2_constraints}. This band has width much less than a percent of the order of magnitude of the axes so is not visible.

In general, indirect detection limits tend to be heavily dependent on the nature of the dark matter distribution, and as such we do not explore them thoroughly in this paper. However, in the case of wino like DM, the limits have been shown to be rather severe~\cite{Cohen_wino_2013,Fan:2013faa}. In particular,  the entire parameter space is excluded by the High Energy Spectroscopic System~\cite{HESS:2013rld} unless the DM distribution shows extremely cored behavior.\footnote{Cored behavior indicates a wide region about the center of the galaxy with a relatively flat DM distribution. One distribution sufficiently cored for wino annihilation to remain undetected by current satellites is a 10 kpc core Burkert profile~\cite{Burkert:1995yz}.} 
Even for such cored models, however, the entire parameter space is expected to be reachable by future CTA measurements~\cite{Cohen_wino_2013}.

\section{Conclusions}

We have shown that split SUSY models with AMSB gaugino masses and thermally produced  DM will be entirely discoverable within the next generations of planned experiments in DM direct detection and electron EDM measurements. The available parameter space is narrow and is of limited scope due to EW symmetry breaking requirements and the consistency with the 125 GeV SM-like Higgs boson mass. If within this framework the DM is composed of Higgsinos, the next generation of experiments will provide a definitive answer on the existence of such new physics. If it is composed of winos, the theory will be discovered or entirely ruled out within the next few generations of electron EDM or indirect detection experiments.

\vspace{12pt}

\noindent
{{\it Acknowledgments}: The work was
supported in part by the DoE DE-SC0007859 (AP, JW), DE-SC0011842 at the University of Minnesota (RC), and
NSF Graduate Research Fellowship (BS).}

\end{document}